\documentclass[a4paper]{jpconf}
\usepackage{graphicx,iopams}
\newcommand{\msun}{M_\odot}

\begin{document}
\title{Multi-band gravitational wave astronomy: science with joint space- and ground-based observations of black hole binaries}

\author{Alberto Sesana}

\address{School of Physics and Astronomy, University of Birmingham, Edgbaston, Birmingham B15 2TT, United Kingdom}

\ead{asesana@star.sr.bham.ac.uk}

\begin{abstract}
  Soon after the observation of the first black hole binary (BHB) by advanced LIGO (aLIGO), GW150914, it was realised that such a massive system would have been observable in the milli-Hz (mHz) band few years prior to coalescence. Operating in the frequency range 0.1-100 mHz, the Laser Interferometer Space Antenna (LISA) can potentially detect up to thousands inspiralling BHBs, based on the coalescence rates inferred from the aLIGO first observing run (O1). The vast majority of them (those emitting at $f<10$ mHz) will experience only a minor frequency drift during LISA lifetime, resulting in signals similar to those emitted by galactic white dwarf binaries. At $f>10$ mHz however, several of them will sweep through the LISA band, eventually producing loud coalescences in the audio-band probed by aLIGO. This contribution reviews the scientific potential of these new class of LISA sources which, in the past few months, has been investigated in several contexts, including multi-messenger and multi-band gravitational wave astronomy, BHB astrophysics, tests of alternative theories of gravity and cosmography. 
\end{abstract}

\section{Introduction}
The detection by advanced LIGO (aLIGO) of a 36$^{+5}_{-4}\msun$ and 29 $^{+4}_{-4}\msun$ coalescing black hole binary (BHB) at $z=0.09^{+0.03}_{-0.04}$ on September 14, 2015 (GW150914) marked the beginning of the era of gravitational wave (GW) astronomy \cite{2016PhRvL.116f1102A}. In the following few months, a second, lighter system (GW151226, \cite{2016PhRvL.116x1103A}), plus a third candidate (LVT151012, \cite{2016PhRvX...6d1015A}) were also observed.

Among the plethora of scientific consequences, those detections made possible for the first time to constrain observationally the coalescence rate of stellar BHBs in the Universe \cite{2016ApJ...833L...1A}. BHBs can in fact originate from a variety of astrophysical channels, including: standard evolution of field massive stellar binaries \cite{2014LRR....17....3P}, dynamical formation in dense stellar clusters \cite{2013LRR....16....4B} or in dense compact remnant cusps surrounding nuclear massive black holes \cite{2016ApJ...831..187A}, evolution of chemically mixed massive binaries \cite{2016MNRAS.458.2634M}, population III remnants \cite{2016MNRAS.460L..74H}, primordial black holes \cite{2016PhRvL.116t1301B}. Every channel, however, is subject to considerable uncertainties. Just to cite the mainstream ones as examples, the number of BHBs resulting from standard stellar evolution critically depends on the poorly understood physics driving common envelope evolution, supernova explosions and their subsequent kicks; similarly, the cluster dynamical formation channel is also affected by the unknown supernova kick velocities and by the retention fraction of BHs in the cluster. The net results is that BHB merger rate predictions vary wildly in the literature (see \cite{2010CQGra..27q3001A} for a review), including some models suggesting rates up to several hundreds events  yr$^{-1}$Gpc$^{-3}$ \cite{2010ApJ...715L.138B,2015ApJ...806..263D}. An extensive analysis of the aLIGO events places the 90\% confidence range between 6 and 200 yr$^{-1}$Gpc$^{-3}$ (depending on the assumed BHB mass distribution \cite{2016PhRvX...6d1015A}), close to the upper end of theoretical predictions. Moreover, aLIGO demonstrated that coalescing BHBs can be in excess of $30\msun$, whereas, based on mass estimates of BHs in X-ray binaries \cite{2011ApJ...741..103F}, it has been often customarily assumed those binaries would have masses around $10\msun$. 

GW150914, in particular, prompted a burst of interest also within the low frequency GW community. Because of its high mass, it was immediately realised that GW150914 would have been visible by a spaceborne interferometer operating in the mHz band few years prior to coalescence \cite{2016PhRvL.116w1102S}. Would the Laser Interferometer Space Antenna (LISA) be operating, we would have known few years in advance GW150914 would have coalesced in the LIGO band on September 14, 2015. Because of the poor constrains on the rates and the standard assumption of $10\msun$ objects, BHBs have never been considered a particularly interesting LISA sources, and multi-band GW observations have been only proposed in the context of seed \cite{2009ApJ...698L.129S} or intermediate mass \cite{2010ApJ...722.1197A} BHBs . On the contrary, analysis of the BHB population as constrained by aLIGO demonstrated that LISA might see up to thousands of these systems, with a variety of profound scientific consequences that are summarized in this contribution. The paper is organized as follows. Section \ref{rates} provides the recipe to construct the BHB population in the LISA band starting from the aLIGO rates and presents detection rates and parameter estimation precision of the observed LISA systems. Section \ref{science} covers several scientific questions that can be tackled using those binaries, and the most relevant points are briefly summarized in Section \ref{outlook}.

\section{Merger rates, observable sources and their properties}
\label{rates}

\subsection{Merger rates}
aLIGO O1 results place the 90\% confidence BHB merger rate within the range 6 -- 200 yr$^{-1}$Gpc$^{-3}$, with most of the uncertainty coming from the unknown shape of the underlying BHB mass function. If the BH mass function is biased-high, then aLIGO observations are consistent with lower intrinsic rates. In \cite{2016ApJ...833L...1A} therefore, rates' probability distributions were computed for two different mass functions: one in which both BH masses, $M_1> M_2$, are drawn from a log-flat mass distribution (model {\it flat}), and one in which the primary BH of a binary is drawn from a Salpeter distribution and the secondary from a flat distribution (model {\it salp}).  The {\it flat} model rate distribution peas at $\approx 35$yr$^{-1}$Gpc$^{-3}$, whereas the {\it salp} model favours relatively light BHBs, with a rate distribution peaking at ($\approx 100$yr$^{-1}$Gpc$^{-3}$).

as detailed in \cite{2016PhRvL.116w1102S}, having assumed a mass distribution, the differential merger rate density $d^2n/d{\cal M}_rdt_r$ can be converted into the number of emitting sources per unit chirp mass, redshift and frequency populating the sky {\it at any time} via 
\begin{equation}
  \frac{d^3N}{d{\cal M}_rdzdf_r}=\frac{d^2n}{d{\cal M}_rdt_r}\frac{dV}{dz}\frac{dt_r}{df_r},
  \label{eq1}
\end{equation}
where ${\cal M}=(M_1M_2)^{3/5}/(M_1+M_2)^{1/5}$ is the chirp mass, $dV/dz$ is the standard volume shell per unit redshift in the fiducial $\Lambda$CDM cosmology ($h=0.679,\,\Omega_M=0.306,\,\Omega_\Lambda=0.694$, \cite{2016A&A...594A..13P}), and $dt_r/df_r\propto f_r^{11/3}$ is given by the standard quadrupole GW emission formula. Subscripts $_r$ refer to quantities evaluated in the source rest frame (as opposed to the detector frame): $t_r=t/(1+z)$, $f_r=(1+z)f$, $M_r=M/(1+z)$. Equation (\ref{eq1}) can be used to generate Monte Carlo realizations of the BHB cosmic population emitting in the LISA band.

\subsection{LISA sensitivity}

\begin{figure*}[h]
  \begin{center}
  \includegraphics[width=30pc]{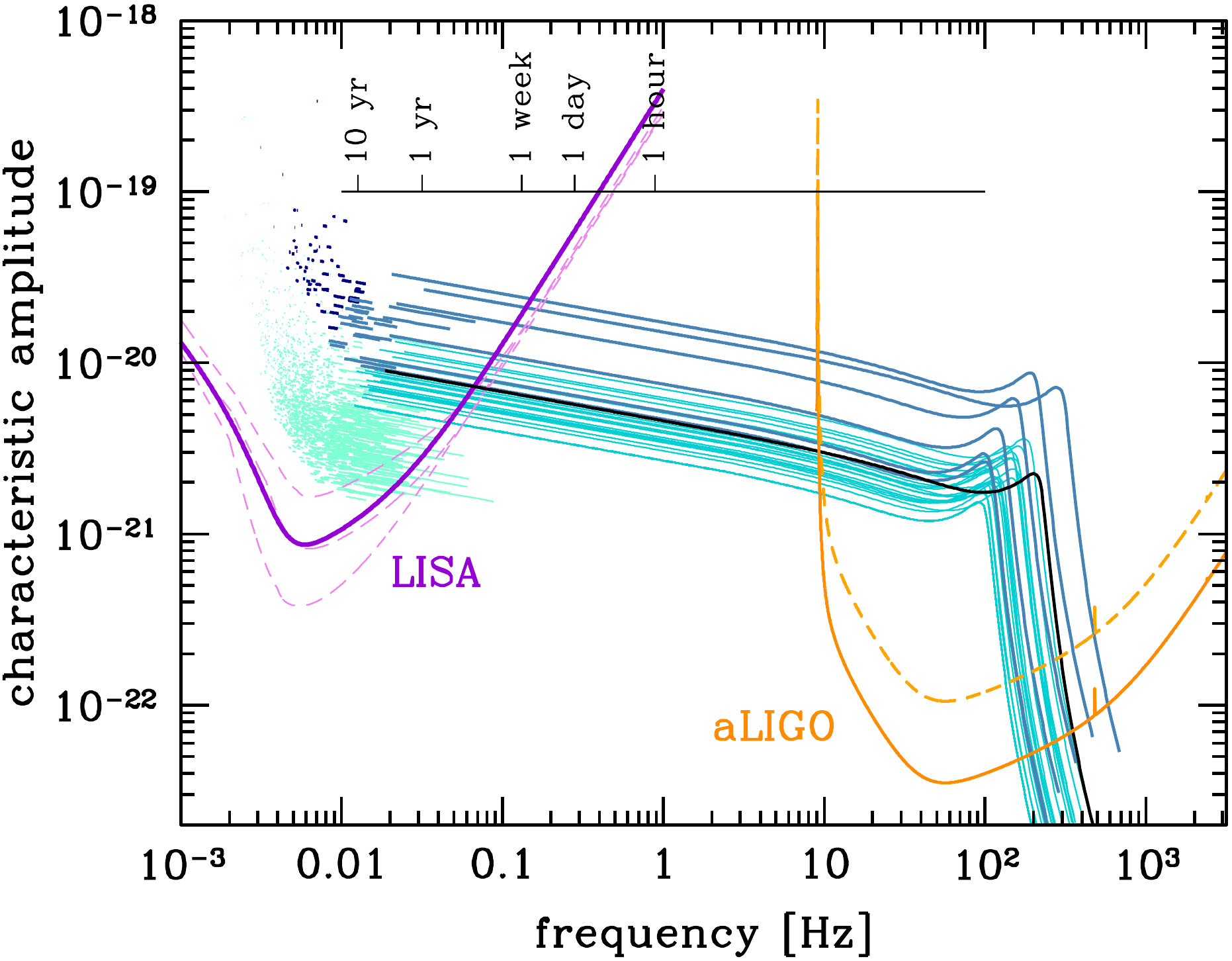}
  \caption{\label{fig1} Visual representation of the multi-band GW astronomy concept. Violet dashed lines are, from top to bottom, the total sensitivity curves of LISA configurations N2A1, N2A2, N2A5 (from \cite{2016PhRvD..93b4003K}). The thick solid purple line is the LISA baseline proposed by the LISA Consortium to address ESA's L3 call. Orange lines refer to current (dashed) and design (solid) aLIGO sensitivity curves. Lines in different shades of blue represent amplitude tracks of BHBs found in a selected Monte Carlo realization of the {\it flat} population model (see main text) seen with S/N$>1$ in the new LISA configuration, integrated assuming a four year mission lifetime (baseline LISA4yr). Light and dark blue curves starting around 0.01Hz and extending to $\sim100$Hz are BHBs coalescing within the LIGO band during the LISA lifetime, and observable by LISA  with S/N$>5$ and S/N$>8$ respectively; the dark blue ticks in the upper left corner are further sources with S/N$>8$ by LISA but not crossing to the aLIGO band within the mission lifetime. Light turquoise lines clustering at the bottom are sources seen in LISA with S/N$<5$ (for clarity those were down-sampled by a factor of 20 and sources extending to the aLIGO band were removed). The characteristic amplitude track completed by GW150914 is shown as a black solid line for comparison. The chart at the top of the figure indicates the frequency progression of this particular source in the last 10 years before coalescence. Adapted from \cite{2016PhRvL.116w1102S}.}
  \end{center}
 \end{figure*}

The LISA concept has significantly evolved in the past few years \cite{AmaroSeoane:2012km,2013arXiv1305.5720C}. In 2014, the European Space Agency (ESA) appointed a gravitational observatory advisory committee (GOAT) to issue a recommendation for a space based GW observatory. During the study, different baselines for a spaceborne interferometer were considered. Full details can be found in \cite{2016PhRvD..93b4003K}. In the following, six baselines featuring one, two or five million km arm-length (A1, A2, A5) and two possible low frequency noises -- namely the LISA Pathfinder goal (N1) and the original LISA requirement (N2) -- are considered. Those configurations, labelled N1A1, N1A2, N1A5, N2A1, N2A2, N2A5, all assume five years of observations with two equivalent Michelson interferometers (i.e. six active laser links). Building on the GOAT recommendation, the LISA Consortium proposed a somewhat different baseline \cite{2017arXiv170200786A}, with slightly different technical specifications, including an armlength of 2.5M kilometres and a mission lifetime requirement of 4 years, and an extension goal to 10 years. The low frequency noise has been set to the level successfully demonstrated by the LISA Pathfinder \cite{2016PhRvL.116w1101A}. We refer to these two latter designs as LISA4yr and LISA10yr. Numbers of detected BHBs are presented below for each of the eight configurations just described.

\subsection{Observed systems and their properties}

\begin{figure}[h]
\includegraphics[width=22pc]{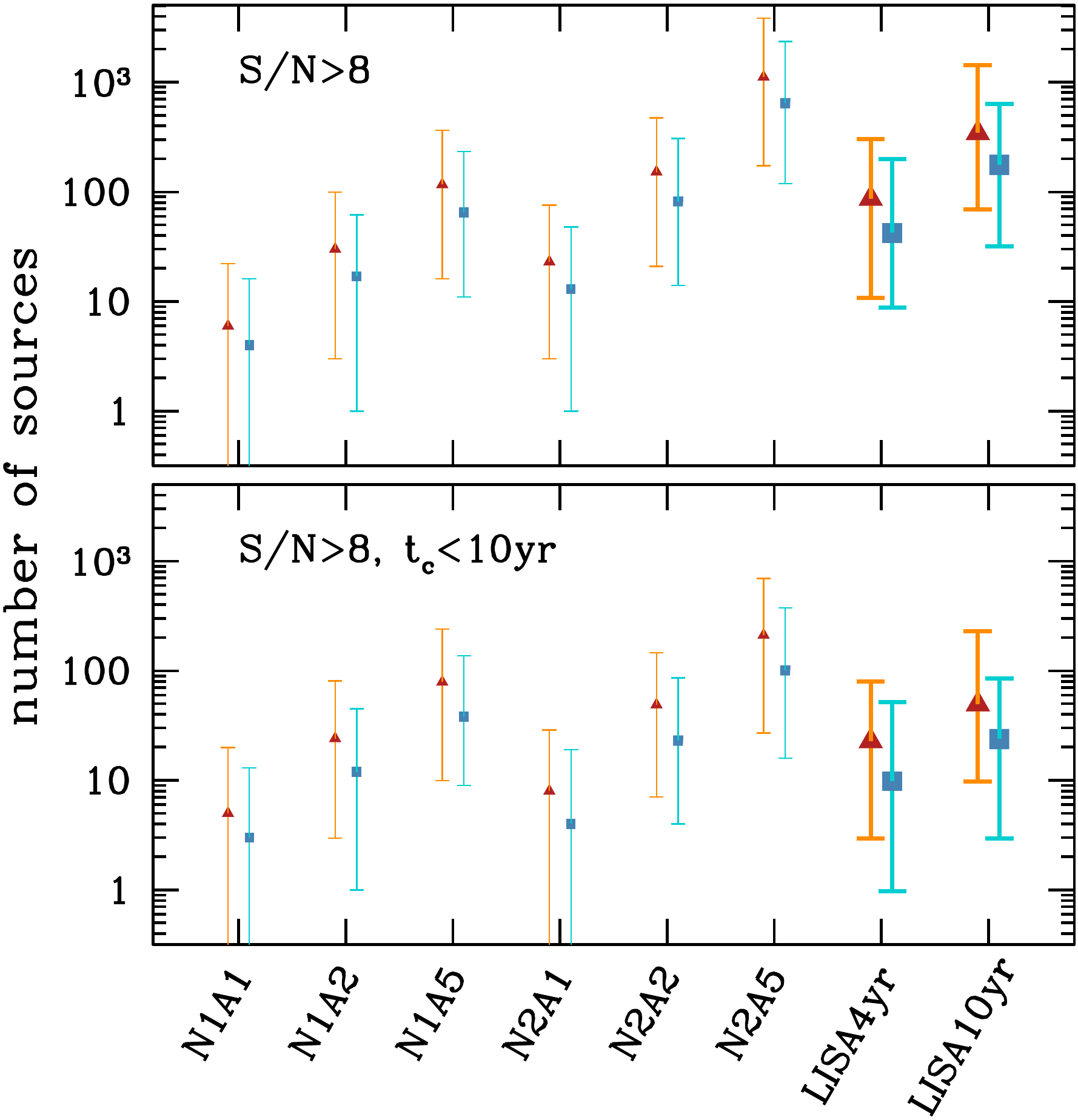}\hspace{2pc}%
\begin{minipage}[b]{14pc}\caption{\label{fig2} Number of resolved BHBs for different LISA baselines (as labelled on the $x$-axis). Orange triangles and blue squares are for models {\it flat} and {\it salp} respectively. Symbols and error-bars are the median and 95\% confidence interval from 200 realizations of the BHB population. The top panel show the total number of sources with S/N$>8$ across the whole LISA band, whereas the lower panel is restricted to sources that will eventually coalesce in the aLIGO band within 10 years from the start of the LISA mission. Results for the LISA configuration proposed to ESA, assuming a mission lifetime of either 4 or 10 years (LISA4yr, LISA10yr) are shown with thicker symbols.}
\end{minipage}
\end{figure}

Depending on the baseline, the number of sources that can be detected above the nominal signal-to-noise ratio (S/N) threshold of eight varies by more than two orders of magnitudes, from just a few in the N1A1 configuration to about a thousand in the N2A5 configuration. Perhaps counter-intuitively, the armlength appear to have a strong impact on the number of detections. This is because, besides shifting the low frequency sensitivity, armlength also severely affects the depth of the bucket, where most of the resolvable BHBs live (see sensitivity curves in figure \ref{fig1}). Therefore, even with equal high frequency noise (as it is the case for all NxAy configurations), the number of detectable sources increases by about two orders of magnitude going from one (A1) to five (A5) million km. LISA4yr performs rather similarly to N2A2, as expected from the similar mission specifications. The sensitivity in the bucket of the two configurations is essentially identical, resulting in a comparable number of total detections with S/N$>8$ (upper panel). Numbers are actually slightly lower for the LISA4yr configuration, because the mission lifetime was assumed to be 5 years in the N2A2 case, and S/N$\propto T^{1/2}$. There is, however, a noticeable difference in the number of sources crossing to the aLIGO band in less then 10 years (lower panel). This is because those accumulate all their S/N at $f>10$ mHz, where the LISA4yr sensitivity becomes about 30-40\% worse than the N2A2 one. Since the cumulative number of sources is proportional to (S/N)$^{3}$, this results in a difference of a factor of more than two in detected systems. LISA would definitely benefit from reaching the 10 year lifetime goal. The number of observable sources is in fact boosted by almost a factor of four in the LISA10yr case, with typical detection numbers of several hundreds. This is again because S/N$\propto T^{1/2}$ and the the cumulative number of sources is proportional to (S/N)$^{3}$. In general the {\rm flat} model, in virtue of its biased-heavy mass function, results in a factor of $\approx 2$ more detections in all cases. Note that for each fixed baseline and BHB population, the 95\% confidence region still spans more than one order of magnitude, because of the intrinsic uncertainties on the aLIGO inferred merger rates. In summary, the current proposed LISA baseline (in its LISA4yr and LISA10yr incarnations) has the potential to observe between few tens to a thousand stellar BHBs with S/N$>8$, and about 10-to-20\% of them will cross to the aLIGO band within 10 years of LISA operation kick-off, realizing the promise of multi-band GW astronomy. 

\begin{figure*}
\centering
\begin{tabular}{cc}
  \includegraphics[width=7.0cm,clip=true,angle=0]{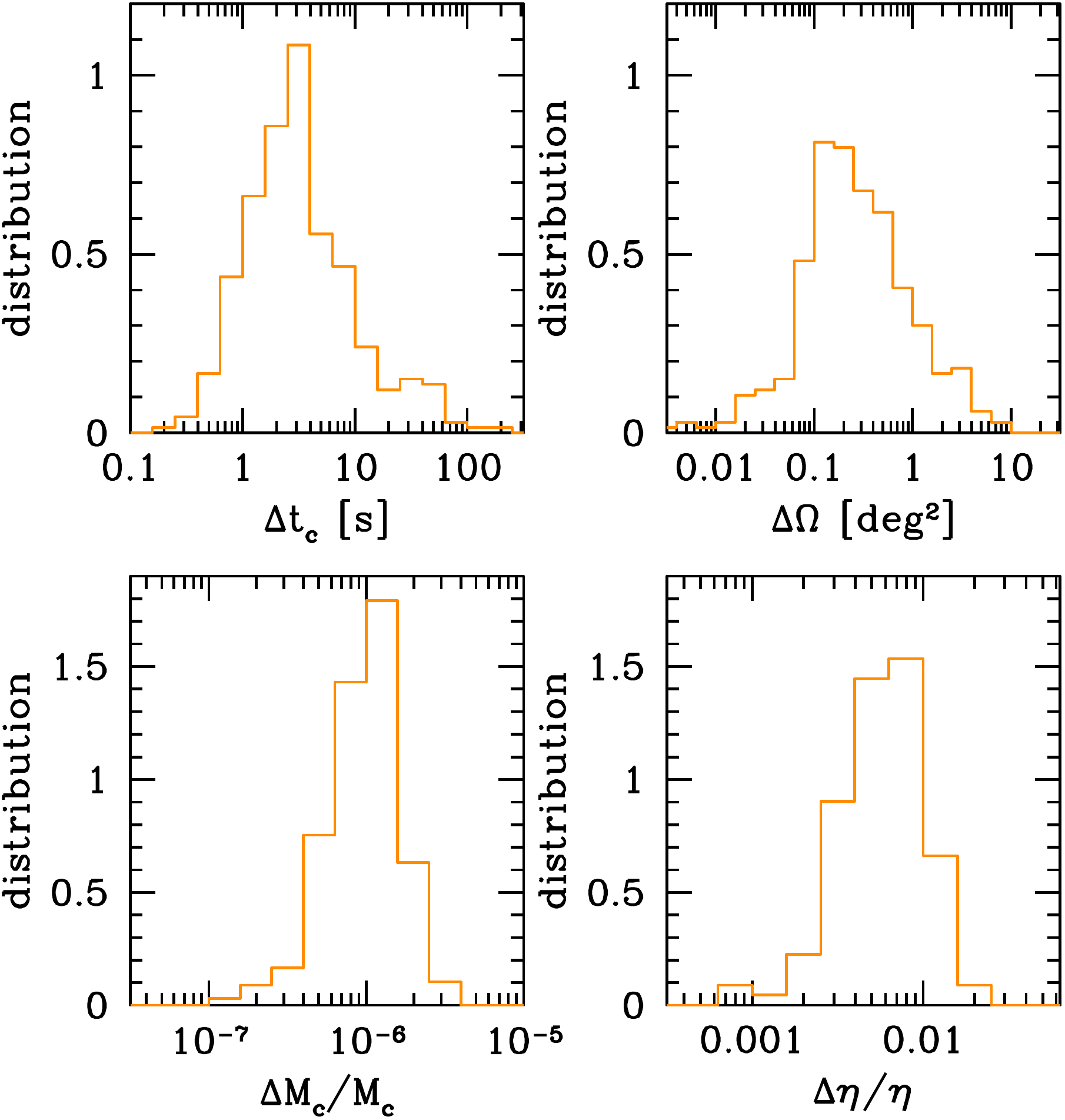}&
  \includegraphics[width=7.0cm,clip=true,angle=0]{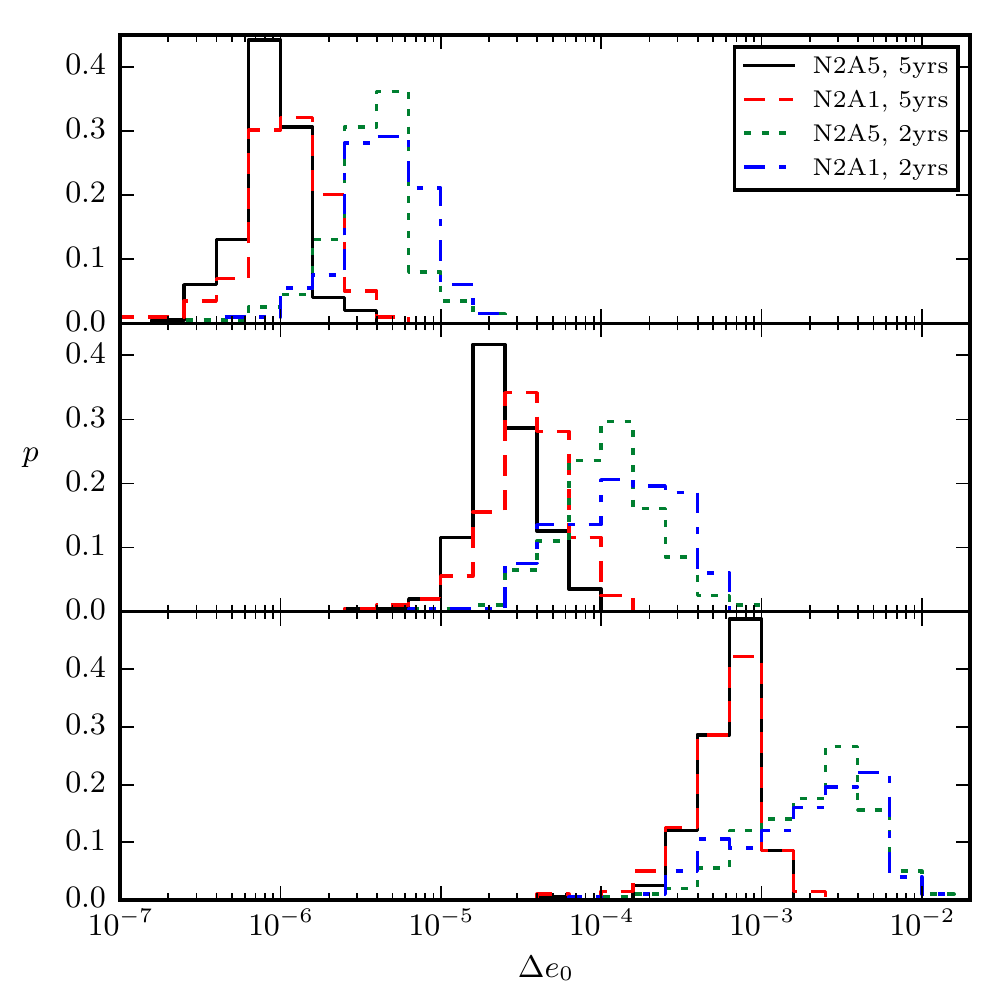}\\
\end{tabular}
\caption{Left figure (from \cite{2016PhRvL.116w1102S}): parameter estimation precision from LISA observations of circular BHBs. Top left: coalescence time; top right: sky localization; bottom left: relative error in the chirp mass ${\cal M}$; bottom right: relative error in the symmetric mass ratio ${\eta}=M_1M_2/(M_1+M_2)^2$. Histograms show normalized distributions obtained from a Monte Carlo realization of 1000 sources observed with S/N$>8$ in the N2A5 configuration, assuming five years of mission operation (results for other configurations look similar). Estimates were obtained via Fisher Matrix analysis using 3.5 Post Newtonian non spinning waveforms \cite{2005PhRvD..71h4025B} and the full time-dependent LISA response function. Right figure (from \cite{2016PhRvD..94f4020N}): errors on the eccentricity measurement at frequency $f_0=10^{-2}$~Hz using eccentric non-spinning waveforms. From top to bottom, the three panels refer to systems with $e_0 = 0.1$, $0.01$ and $0.001$. Linestyles refer to different LISA baselines: N2A5 and $T_{\rm obs}=5\,{\rm yrs}$ (solid black), N2A1 and $T_{\rm obs}=5\,{\rm yrs}$ (dashed red), N2A5 and $T_{\rm obs}=2\,{\rm yrs}$ (dotted green), N2A1 and $T_{\rm obs}=2\,{\rm yrs}$ (dash-dotted blue).}
\label{fig3}
\end{figure*}


Early calculations by \cite{2016PhRvL.116w1102S} also showed that LISA can achieve exquisite accuracy in the measurement of selected BHB parameters. This is shown in figure \ref{fig3} for systems that {\it cross to the aLIGO band} within the LISA lifetime. Time to coalescence can be typically established within less than 10 seconds and sky location accuracy is generally better than a square degree. The redshifted chirp mass and symmetric mass ratio can be determined to a $10^{-6}$ and 0.01 relative accuracy respectively (which means that individual redshifted masses can be measured to better than 1\%). Note that for these sources, LISA and aLIGO measurement can be combined to reduce parameter estimation errors \cite{2016PhRvL.117e1102V}. Considering mildly eccentric (up to 0.1) systems, \cite{2016PhRvD..94f4020N} showed in a follow-up paper that eccentricities in excess of $10^{-3}$ can be confidently measured, as shown in the right panel of figure \ref{fig3}, which can provide useful information on the BHB formation channel (see below). Other parameters might not be determined with comparable precision. For example, the distance to the source can only be determined to $\Delta{D_l}/D_l\approx A/({\rm S/N})$, where in general $A>1$; which gives rather poor results, since most BHBs have S/N$<20$. Also spin determination might be problematic in general, as the two BHs are separated by thousands of Schwarzschild radii in the LISA band and the spin-orbit and spin-spin coupling corrections on the waveform are accordingly small. An extensive study on parameter estimation precision, including precessing spinning waveforms is currently ongoing (Klein et al. in preparation).

\section{Scientific potential}
\label{science}
Because of the relatively large number of observable systems in the LISA band, the exquisite precision to which some of the parameters will be measured, and the possibility of seeing few of them sweeping through the aLIGO band within few years of the initial LISA detection, massive stellar BHBs captured the interest of the wider GW community. This fairly 'unexpected' (in terms of typical mass and rates) class of sources can be exploited in several contexts, including BHB astrophysics, fundamental physics and cosmology. This section focuses on some of the main payouts, keeping in mind that this is relatively new investigation ground, and the list is inevitably incomplete. 

\subsection{Multi-band and multimessenger astronomy}
One of the obvious benefits of sources like GW150914 is the realization of multi-band GW astronomy. Observations of the same signal in two different detectors provides an efficient independent way to cross check and validate the instruments (which is particularly valuable for a space-based detector). Moreover, the two instruments will observe at separate frequency bands, therefore covering different evolutionary stages of BHBs; while LISA will be sensitive to their adiabatic inspiral, most of the aLIGO S/N comes from the very late inspiral, merger and ringdown. Each stage of the evolution carries information about different physical aspects of the sources, that can be combined together to push the boundary of GW astronomy in several directions, as highlighted below.

Another obvious advantage of observing BHBs several months before coalescence, is the possibility of issuing early warnings for an upcoming coalescence. LISA will allow exquisite estimation of some key parameters of the BHBs crossing to the aLIGO band. In particular, the time to coalescence can be predicted within less then ten seconds (but see \cite{2016arXiv160908093B} for potential complications) and the sky localization known within less than a square degree few weeks {\it before} merger. Although it is not standard to associate electromagnetic (EM) counterparts to BHB mergers, several (sometimes exotic) EM production scenarios have been proposed (see e.g.\cite{2016ApJ...821L..18P,2016ApJ...822L...9M,2016ApJ...823L..29K}), triggered by a tentative association of a gamma-ray signal detected by Fermi to GW150914 \cite{2016ApJ...826L...6C}. The claim is controversial, and the signal has not been confirmed, among others, by simultaneous INTEGRAL observations \cite{2016ApJ...820L..36S}. In any case, since the Universe is certainly not short of surprises, it is wise keep an open mind and continue to point our telescopes, and this can be done better if we know in advance when and where to observe. Knowing the sky location will facilitate the early pointing of large field of view instruments at all wavelengths. In particular, deep searches can be performed {\it in coincidence} to the coalescence, testing any possible exotic scenario of emission on the BHB merger dynamical timescale. 

\subsection{BHB astrophysics}

\begin{figure}[h]
\includegraphics[width=18pc]{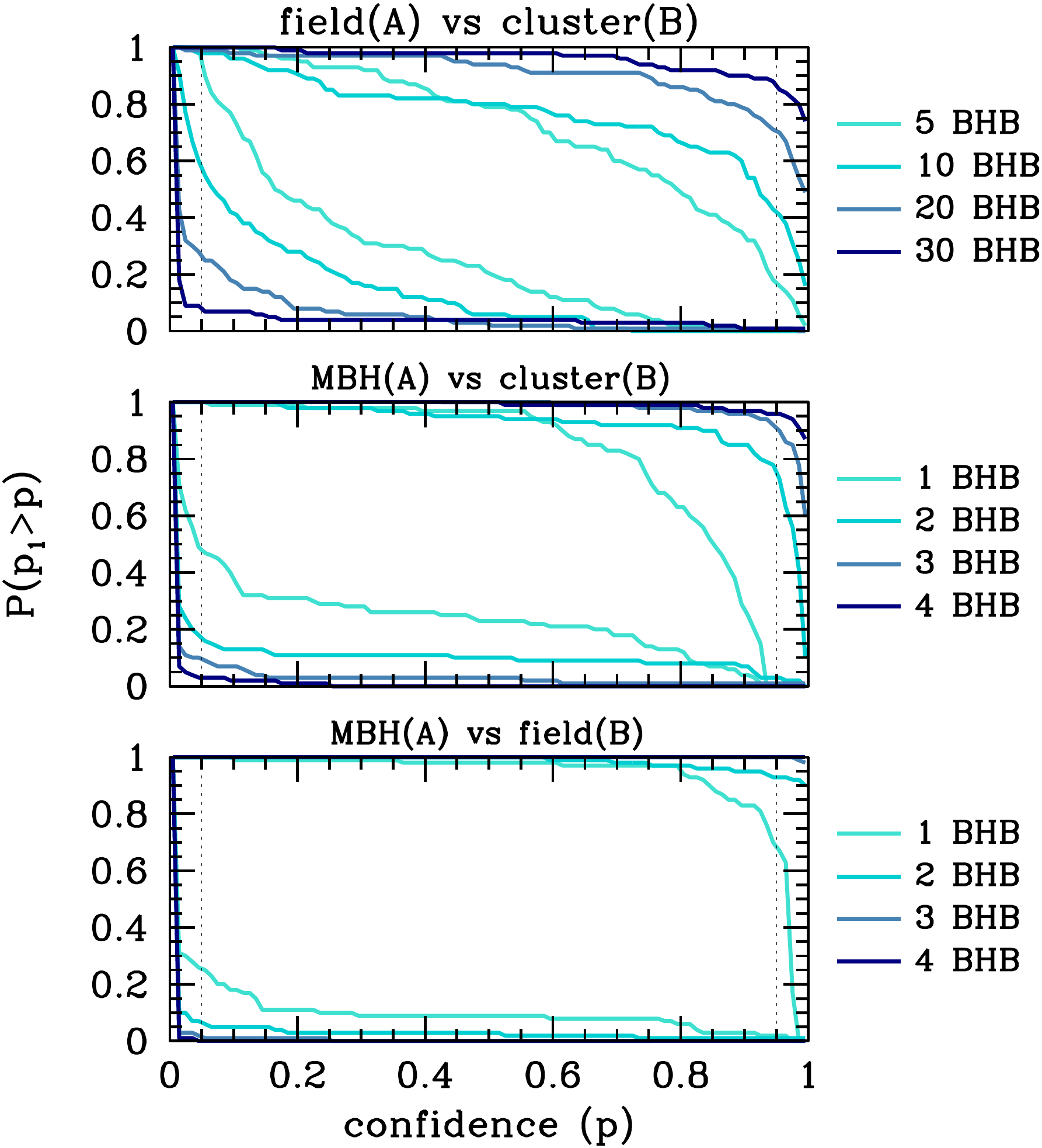}\hspace{1pc}%
\begin{minipage}[b]{19pc}\caption{\label{fig4} Cumulative distribution function (CDF) of the confidence in a given model over 100 Monte Carlo realizations of a specific number, $N_{\rm obs}$, of observed BHBs (several $N_{\rm obs}$ are shown in each panel as labelled on the right). The N2A2 configuration is assumed, but results are weakly dependent on the adopt LISA baseline. For each pair of curves, the top one marks the CDF of confidence in model $A$ when model $A$ is true, whereas the bottom one marks the CDF of confidence in model $A$ when model $B$ is true. For example, the upper panel compares the {\it field} (massive binary evolution) and {\it cluster} (dynamical capture) scenarios; observation of 30 binaries are sufficient to distinguish among them at a 95\% confidence level in more than 90\% of the realizations. From \cite{2017MNRAS.465.4375N}.}
\end{minipage}
\end{figure}

Combining high and low frequency GW detection, will also help in the identification of the astrophysical channel responsible of BHB formation. Different scenarios in fact result in different mass, mass ratio, spin  and eccentricity distribution of the detected sources. Ground and space based detectors are conveniently sensitive to different parameters, providing complementary information. By measuring the last few inspiral cycles with high S/N, a network of second generation ground based detectors will allow to measure individual masses within 10\% precision. Information about spin magnitude and direction can be extracted by observing the effect of spin-orbit coupling induced precession, which is maximum in the late inspiral. However, because of GW circularization, BHBs will generally have a non-measurable eccentricity $<0.01$ in the aLIGO band, regardless of their formation channel. LISA, on the other hand, by observing up to millions GW cycles, will allow the measurement of individual (redshifted) masses to better than 1\% (and often better than 0.1\%). Individual spins can be measured in a fraction of cases (Klein et al in preparation), but generally not to high precision. But perhaps most importantly, being sensitive to lower frequencies, LISA will catch GW signals from BHBs that did not have enough time to fully circularize, measuring any eccentricity in excess of $e\approx 10^{-3}$. As independently pointed out by \cite{2017MNRAS.465.4375N} and \cite{2016ApJ...830L..18B}, this will allow to differentiate between competing formation scenario, chiefly between dynamically formed systems and binary stellar evolution remnants. Figure \ref{fig4}, shows an example of how competitive formation scenarios can be distinguished after a given number of LISA detections, on the basis of eccentricity measurements only. The two lower panels show that a handful of measurements are enough to recognize merger induced by Kozai cycles triggered by the presence of a massive black hole in the proximity of the BHB (model {\it MBH} from \cite{Antonini:2012ad}). Conversely, more than 30 detections are required to efficiently differentiate between dynamical formation (model {\it cluster} from \cite{Rodriguez:2016kxx}) and stellar evolution remnants (model {\it field} from \cite{Kowalska:2010qg}). The currently proposed LISA design will likely detect few hundred such systems. Eccentricity information can also be combined with mass and spin measurements (from both LISA and aLIGO) to optimize astrophysical inference.

\subsection{Tests of alternative theories of gravity}

\begin{figure}[h]
\includegraphics[width=20pc]{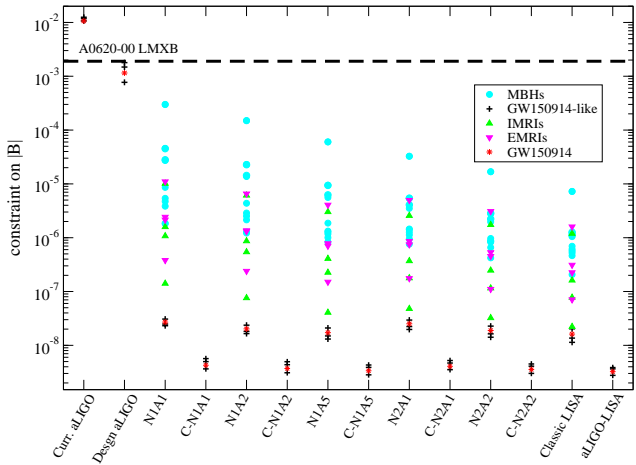}\hspace{1pc}%
\begin{minipage}[b]{17pc}\caption{\label{fig5} $1\sigma$ constraints on the BH dipole flux parameter $B$ as a function of the assumed GW detector from different GW sources: GW150914-like systems (stars and pluses), massive BH binaries (filled circles) and EMRIs/IMRIs (filled triangles). Shown are the NxAy designs proposed in the GOAT study as well as Classic LISA; the prefix C- indicates observations in combination with aLIGO ad design sensitivity. The current best limit on vacuum dipole radiation from LMXB A0620-00~\cite{PhysRevD.86.081504} is shown as a dashed horizontal line for comparison. From \cite{2016PhRvL.116x1104B}.}
\end{minipage}
\end{figure}

Multi-band detections will also enhance the potential of gravity tests in the strong, dynamical field regime of merging BHBs. Massive systems will be observed by ground based detectors with high merger (and possibly ringdown) S/N, after being tracked for years by LISA in their adiabatic inspiral. The two portions of signal can be combined to push the search for deviations from General Relativity, due for example to dipolar radiation, as first discussed in \cite{2016PhRvL.116x1104B}. Although timing of binary pulsars excludes the emission of dipole gravitational radiation by binaries \cite{2010PhRvD..82h2002Y}, entire classes of theories predict this effect predominantly (or only) in binaries involving BHs (see \cite{2015CQGra..32x3001B} for a review). Joint observations of GW150914-like systems by aLIGO and LISA can potentially improve current bounds on dipole emission from BHBs by more than six orders of magnitude, as shown in figure \ref{fig5}. The figure highlights that even if observed by LISA alone, BHBs constrain these theories better than massive BHBs (MBHBs) and extreme mass ratio inspirals (EMRIs). This is because the signal has to be phased for millions of cycles to be recovered, being therefore sensitive to minuscule phasing deviations caused by dipole emission. When combined to LIGO observations of the same system, typical bounds improve by an additional factor of about five. 

\subsection{Cosmography and cosmology}
Stellar BHBs observed in the LISA band potentially provide a new class of standard candles. In absence of a distinctive EM counterpart, \cite{2016arXiv160907142K} estimated the number of BHBs that can be localized with enough precision that only one galaxy falls in the GW measurement error cube. They concluded that LISA might measure $H_0$ to a few\% precision, based on the analytical scaling of the average $\Delta{\Omega}$ and $\Delta{D_l}/D_l$ and the average number density of possible host galaxies. Although this is an useful result, it relies on average estimates of quantities ($\Delta{\Omega}$ and $\Delta{D_l}/D_l$) that vary by orders of magnitude from one source to another. Moreover, the outcome depends severely on the uncertain BHB merger rate and on the LISA design. A systematic investigation is currently ongoing (del Pozzo et al., in preparation). Preliminary results show in fact that  $\Delta{D_l}/D_l$ errors in \cite{2016arXiv160907142K} are generally underestimated and association with an individual galaxy is not possible. One has therefore to rely on building a likelihood function based on the redshift distributions of plausible galaxy hosts in the GW measurement error cube. The technique has been first applied by \cite{2008PhRvD..77d3512M} to EMRIs and subsequently by \cite{2011ApJ...732...82P} to MBHBs. Note that LISA will provide excellent sky localization up to $\Delta{\Omega}<0.1$deg$^2$ precision, it will not, however, provide a very good measurement of the luminosity distance, $D_l$, to the source. This latter might in fact be better measured by ground based detectors, that might see those sources with S/N$>100$ at design sensitivity. Multi-band BHBs crossing between the two detectors' windows, might therefore be particularly valuable standard candles, combining LISA sky localization with aLIGO distance determination.

\subsection{Milky way BHBs}
As first pointed out in \cite{2016MNRAS.460L...1S} and then further elaborated in \cite{2017arXiv170101736C}, LISA might also detect few such BHBs within the Milky Way. This specific prospect is particularly appealing for a number of reasons. Considering the typical LISA error-cube, the location of those system within the Milky Way will be rather accurate, shedding light on their formation mechanism. In case of few detections, a consistent presence of star clusters or globular clusters within the sky localization error would be a decisive element in support of the dynamical formation scenario (viceversa, the lack thereof would not discard this model, since BHBs can be efficiently ejected from their parent clusters, and the clusters themselves can evaporate by the time the BHBs merge). A frequent localization towards the Galactic centre might provide evidence of the role of dense BH cusps surrounding massive black holes in the formation of these systems, as predicted by some formation channels (see e.g. \cite{2017ApJ...835..165B}). Their proximity will also allow to perform deep searches for associated EM signals. If BHBs are surrounded by debris disks left behind by partially failed supernovae, even a small amount of accretion leads to an observable photon flux within galactic distances. In the unlikely event that many such BHBs are detected, then their mass function can be linked to the metallicity content of the MW, to gather further clues about the physics driving their formation. 

\section{Conclusions}
\label{outlook}
The first aLIGO detections showed that stellar BHBs with chirp masses in excess of $\approx20\msun$ are more abundant than previously thought, and the proposed LISA has the potential to see up to about a thousand of them, depending on the adopted mission design. About 10\% of these systems, those caught emitting at $f>10$ mHz, will sweep through the LISA window, eventually coalescing in the audio band probed by aLIGO. Multi-band and multi instrument observations of these sources will open a number of new possibilities for GW astronomy. Eccentricity measurement of BHB in the LISA band will allow to disentangle between different formation channels, and their exquisite sky localization will provide an independent local measurement of the Hubble constant. The phasing of millions of waveform cycles will constrain theories of gravity allowing dynamical scalarization with unprecedented precision, especially if combined with aLIGO observations of the merger phase. In general, joint and complementary LISA-aLIGO observations of the same source will enhance the scientific potential by combining measurements of the early inspiral (LISA) to observations of merger and ringdown (aLIGO). Moreover, S/N accumulated over years of observations in the LISA band will allow to issue early warnings (with few weeks notice) about when and where the BHB will coalesce in the LIGO band. Ground and space based probes across the EM spectrum can therefore be pre-pointed in the direction of the system, performing a deep search for possible EM precursors or for bursts coincident with the final coalescence. Multi-band GW observations of BHBs add a new exciting goal to the already rich LISA science case, opening new avenues in multi-messenger astronomy.     

\section*{References}
\bibliography{bibnote}

\end{document}